\documentclass[epj]{svjour}
\usepackage{amssymb,amsmath}
\usepackage{graphicx}
\begin{document}
\title{Long-lived states of oscillator chain with dynamical traps}

\author{Ihor Lubashevsky\inst{1} \and Reinhard Mahnke\inst{2}
 \and Morteza Hajimahmoodzadeh\inst{3}\and
Albert Katsnelson\inst{3}
}                     
%
%
\institute{General Physics Institute named after A.M. Prohorov, Russian Academy of Sciences,\\
           Vavilov str., 38, 119991 Moscow, Russia, e-mail: ialub@fpl.gpi.ru
\and
           Fachbereich Physik, Universit\"at Rostock, \\D 18051 Rostock, Germany,
           e-mail: reinhard.mahnke@physik.uni-rostock.de
\and
       Faculty of Physics, M.V. Lomonosov Moscow State University, \\ Moscow 119992, Russia,
           e-mail: albert@solst.phys.msu.su}
\date{Received: date / Revised version: date}
%
\abstract{ A simple model of oscillator chain with dynamical traps and additive
white noise is considered. Its dynamics was studied numerically. As
demonstrated, when the trap effect is pronounced nonequilibrium phase
transitions of a new type arise. Locally they manifest themselves via
distortion of the particle arrangement symmetry. Depending on the system
parameters the particle arrangement is characterized by the corresponding
distributions taking either a bimodal form, or twoscale one, or unimodal
onescale form which, however, deviates substantially from the Gaussian
distribution. The individual particle velocities exhibit also a number of
anomalies, in particular, their distribution can be extremely wide or take a
quasi-cusp form. A large number of different cooperative structures and
superstructures made of these formations are found in the visualized time
patterns. Their evolution is, in some sense, independent of the individual
particle dynamics, enabling us to regard them as dynamical phases.
} 
\maketitle

\section{Introduction\label{intro}}

For the last several decades various phenomena cause by the ordering action of
noise in nonequilibrium systems are found (for a general review see
Refs.~\cite{B1,B2,B3,B4}). Popular examples are stochastic
resonance~\cite{Str1,Str2,Str3}, coherence resonance~\cite{B2,Cohr},
noise-induced transport~\cite{NIT}, and noise-induced phase
transitions~\cite{B3,NIPT1,NIPT2}. Typically the latter are due to
multiplicative noise. However, additive noise in the presence of other
multiplicative noise can also induce phase transitions~\cite{Z1,Z2,Z3} or
individually cause a hidden phase transition to become visible~\cite{Z4}.

The constructive role of noise is peculiar to nonequilibrium systems only. In
thermodynamic systems, for example, the phase formation is solely due to a
certain regular ``force'' changing its form, in particular, the number of
stationary points. Available noise mainly perturbs the system motion around
these points.

In the present paper we pay attention to a new class of nonequilibrium systems,
namely, many particle ensembles where some long-lived cooperative states can
form whereas the regular component of ``individual'' forces has no stationary
points except for one corresponding to the homogeneous state. The latter, in
addition, is locally stable for all the values of the system parameters.
Besides, only additive noise enters such systems.

Originally investigation of the model under consideration was stimulated by a
wide class of intricate cooperative phenomena found in the dynamics of vehicle
ensembles moving on highways, motion of fish and bird swarms, stock markets,
\textit{etc} (for a review see Ref.~\cite{Hel}). The model background is the
following. People as elements of a certain system cannot individually control
all the governing parameters. Therefore one chooses a few crucial parameters
and focuses on them the main attention. When the equilibrium with respect to
these crucial parameters is attained the human activity slows down retarding,
in turn, the system dynamics as a whole. For example, in driving a car the
control over the relative velocity is of prime importance in comparison with
the correction of the headway distance. So under normal conditions a driver,
first, should eliminate the relative velocity between his car and a car ahead
and only then optimize the headway.

These speculations have led us to the concept of dynamical traps, a certain
``low'' dimensional region in the phase space where the main kinetic
coefficients specifying the time scales of the system dynamics become
sufficiently large in comparison with their values outside the trap region
\cite{we1,we2,we3,we4}. As a result long-lived states have to appear. In time
patterns these states manifest themselves like a sequence of fragments within
which at least one of the phase variables remains approximately constant. These
fragments are continuously connected by sharp jumps of the given variable.
Paper~\cite{we2} demonstrated that such long-lived states do exist in the dense
traffic flow and proposed some model of dynamical traps to explain the observed
features of car velocity time series. Papers~\cite{we3,we4} simplified this
model to single out the dynamical trap effect in its own record. Similar
phenomena seem to be observed in physical systems also, for example, during the
non-monotonic relaxation of Pd-metal alloys charged with hydrogen \cite{Kat1}.
Paper~\cite{we3} studied a single oscillator with dynamical traps and
demonstrated numerically that white noise can cause the distribution function
of oscillator position to convert from the unimodal form to the bimodal one. It
is due to the fact that inside the trap region the regular ``force'' is
depressed only rather than changes the sign and the system motion is mainly
caused by a random Langevin ``force''. A first step towards of this effect in
oscillator ensembles was made in Ref.~\cite{we4}. In particular, the dynamical
traps were demonstrated to be able to give rise to the system instability and
an anomalous velocity distribution like a cusp $\propto \exp\{-|v|\}$ smoothed,
naturally, inside a narrow transition region. It should be noted that similar
anomalous velocity distributions were found for dense traffic flow \cite{Wag}.

The purpose of the present paper is to demonstrate that an ensemble of such
oscillators with dynamical traps can exhibit a number of anomalous cooperative
phenomena, their detailed investigation will be published elsewhere. However,
first, we clarify the relation between the specific mathematical form of the
model to be studied and the concept of dynamical traps discussed above.

\subsection*{Motivated behavior of particles}

Keeping in mind the aforesaid about the human behavior we consider a
one-dimensional ensemble of ``lazy'' particles characterized by their positions
and velocities $\{x_i,v_i\}$ as well as possessing some motives for active
behavior. Particle~$i$ wants to get the ``optimal'' middle position between the
nearest neighbors. So one of the stimuli for it to accelerate or decelerate is
the difference $\eta_i = x_i - \frac12(x_{i-1} + x_{i+1})$ provided its
relative velocity $\vartheta_i = v_i -\frac12(v_{i-1} + v_{i+1})$ with respect
to the pair of the nearest neighbors is sufficiently low. Otherwise, especially
if particle $i$ is currently located near the optimal position, it has to
eliminate the relative velocity, being the other stimulus for particle $i$ to
change its state of motion. Since a particle cannot predict the dynamics of its
neighbors it has to regard them as moving uniformly with the current
velocities. So both the stimuli are to determine directly its acceleration
$dv_i/dt$. The model to be formulated in the next section combines both of
these stimuli within a linear approximation  similar to $(\eta_i +
\sigma\vartheta_i)$, where $\sigma$ is the relative weight of the second
stimulus.

When, however, the relative velocity $\vartheta_i$ of particle $i$ attains
sufficiently low values the current situation for it cannot become worse, at
least, rather fast. So in this case particle $i$ prefers not to change the
state of motion and to retard the correction of its relative position. This
assumption leads to the appearance of some common cofactor
$\Omega(\vartheta_i)$ in the governing equation like this
$$
   \frac{dv_i}{dt} \propto
   - \Omega(\vartheta_i)\big(\eta_i + \sigma\vartheta_i \big)\,.
$$
The cofactor $\Omega(\vartheta)$ is to meet the inequality $\Omega(\vartheta)\ll 1$ for $\vartheta\ll\vartheta_c$
and $\Omega(\vartheta)\approx 1$ when $\vartheta\gg \vartheta_c$, where $\vartheta_c$ is a certain critical value
quantifying the particle perception of speed. Exactly the appearance of such a
factor is the implementation of the dynamical trap effect.
Now let us specify the model.

\section{Model\label{sec:1}}

\begin{figure}
 \begin{center}
 \includegraphics[width = 0.9\columnwidth]{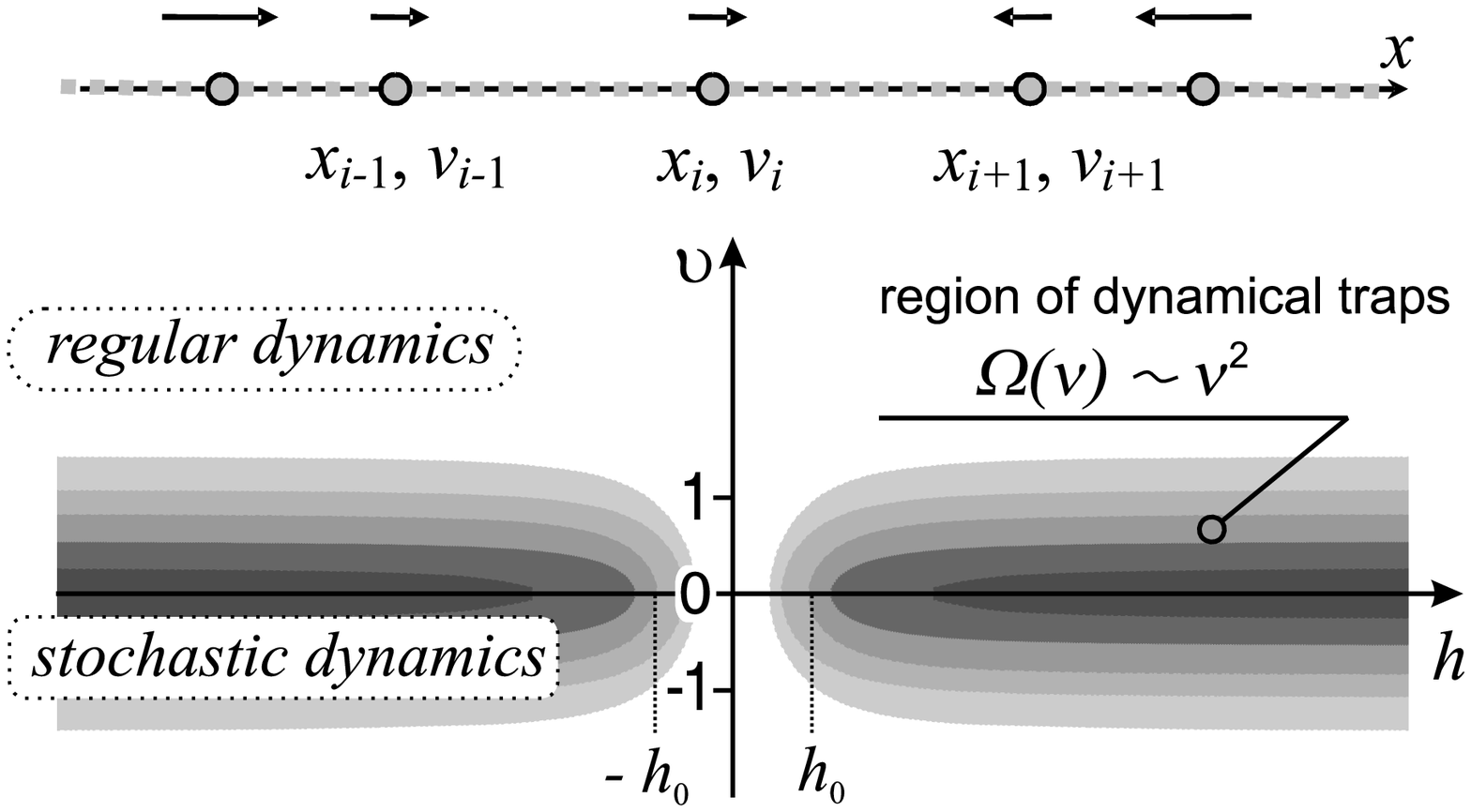}
 \end{center}
\caption{
  The particle ensemble under consideration and the structure of the phase
  space. The darkened region depicts the points where the dynamical trap effect
  is pronounced. For the relationship between the variables $x_i$, $v_i$,
  $h_i$, and $\vartheta_i$ see formulae~\eqref{4} and \eqref{4a}.
  \label{F1}}
\end{figure}

The following linear chain of $N$ point-like particles is considered
(Fig.~\ref{F1}). Each internal particle $i\neq 1, N$ can freely  move along the
$x$-axis interacting with the nearest neighbors, namely,  particles $i-1$ and
$i+1$ via ideal elastic springs with some quasi-viscous friction. The dynamics
of this particle ensemble is governed by the collection of coupled equations
\begin{align}
  \label{1}
  \frac{dx_i}{dt} & = v_i\,,\\
  \label{2}
  \frac{dv_i}{dt} & = - \Omega(\vartheta_i,h_i)[\eta_i + \sigma \vartheta_i + \sigma_0 v_i]
   + \epsilon \xi_i(t)\,.\\
\intertext{Here for $i = 2,3,\ldots,N-1$ the variables $\eta_i$ and
$\vartheta_i$ to be called the symmetry distortion and the distortion rate,
respectively, are specified as}
 \label{3}
 \eta_i & = x_i - \frac12(x_{i-1}+x_{i+1})\,, \\
 \label{4}
 \vartheta_i & = v_i - \frac12(v_{i-1}+v_{i+1})\,,\\
\intertext{the mean distance $h_i$ between the particles at the point $x_i$, by
definition, is} \label{4a} h_i & = \frac12(x_{i+1} - x_{i-1})\,,
\end{align}
and $\{\xi_i(t)\}$ is the collection of mutually independent  white noise
sources of unit amplitude, i.e.
\begin{equation}\label{5}
   \big\langle \xi_i(t) \big\rangle = 0\,, \quad
   \big\langle \xi_i(t)\xi_{i'}(t') \big\rangle = \delta_{ii'}\delta(t-t')\,.
\end{equation}
In addition, the parameter $\epsilon$ is the noise amplitude, $\sigma$ is the
viscous friction coefficient of the springs, $\sigma_0$ is a small parameter
that can be treated as a certain viscous friction related to the particle
motion with respect to the given physical frame. It is introduced to prevent
the system motion as a whole with infinitely high velocity. Besides, the symbol
$\langle\ldots \rangle$ denotes averaging over all the noise realizations,
$\delta_{ii'}$ and $\delta(t-t')$ are the Kronecker symbol and the Dirac
$\delta$-function. The factor $\Omega(\vartheta_i,h_i)$ is due to the effect of
dynamical traps and actually following our previous paper~\cite{we1} the
\textit{Ansatz}
\begin{align}
 \label{6}
   \Omega(\vartheta,h) & = \frac{\vartheta^2 + \triangle^2(h)}{\vartheta^2 + 1}\\
\intertext{with a function $\triangle(h)$ such that}
 \label{6a}
 \triangle^2(h) & = \triangle^2 +\big(1-\triangle^2\big) \frac{h_0^2}{h^2 + h_0^2}
\end{align}
is used. The parameter $\triangle\in [0.1]$ quantifies the dynamical trap
influence and the spatial scale $h_0$ specifies the small distances within
which the trap effect is to be depressed, i.e. for $h\ll h_0$ the value
$\triangle(h)\approx 1$ whereas when $h\gg h_0/\triangle$ the value
$\triangle(h) \approx \triangle$. If the parameter $\triangle = 1$, the
dynamical traps do not exist at all, in the opposite case, $\triangle\ll 1$,
their influence is pronounced inside a certain neighborhood of the $h$-axis
(trap region) whose thickness is about unity (Fig.~\ref{F1}). The temporal and
spatial scales have been chosen so that the thickness of the trap region be
about unity as well as the oscillation circular frequency be also equal to
unity outside the trap region. The terminal particles, $i = 1$ and $i = N$, are
assumed to be fixed, i.e.
%
\begin{equation}\label{7}
 x_1(t)  = 0\,, \qquad x_N(t)  = (N-1)l\,,
\end{equation}
where $l$ is the particle spacing in the homogeneous chain. The particles are
treated as mutually impermeable ones. So when the coordinate $x_i$ and
$x_{i+1}$ of an internal particle pair become identical the absolutely elastic
collision is assumed to happen, i.e. if $x_i(t) = x_{i+i}(t)$ at a certain time
$t$ then the timeless velocity exchange
\begin{equation}\label{8a}
  \begin{aligned}
   v_i(t+0) & = v_{i+1}(t-0) \,, \\ v_{i+1}(t+0) & = v_i(t-0)
  \end{aligned}
\end{equation}
comes into being.  The multiparticle collisions are ignored.

The system of equations~\eqref{1}--\eqref{8a} forms the model under
consideration.

The stationary point $x^\text{st}_i = (i-1)l$ is stable with respect to small
perturbations. It steams from the linear stability analysis with respect to
perturbations of the form
\begin{equation}\label{9}
 \delta x_i(t)\propto \exp\{\gamma t + \mathbf{i}kl(i-1)\}\,,
\end{equation}
where $\gamma$ is the instability increment, $k$ is the wave number, and the
symbol $\mathbf{i}$ denotes the imaginary unite. The boundary
conditions~\eqref{7} are fulfilled by assuming the wave number $k$ to
take the values $k_m = \pi m/[(N-1)l]$ for $m = \pm1,\pm2,\ldots,\pm(N-2)$. For
large values of the particle number $N$ the parameter $k$ can be treated as a
continuous variable. Using the standard technique the system of
equations~\eqref{1}, \eqref{2} for perturbation~\eqref{9} leads us to the
following relation of the instability increment $\gamma (k)$ and the wave
number $k$:
\begin{multline}
  \gamma = - \Omega_0\bigg[ \frac12\sigma_0+  \sigma \sin^2\Big(\frac{kl}2 \Big) \bigg] \\
  {}+\mathbf{i}\sqrt{2\Omega_0 \sin^2\Big(\frac{kl}2 \Big) -
  \Omega_0^2
  \bigg[ \frac12\sigma_0+  \sigma \sin^2\Big(\frac{kl}2 \Big) \bigg]^2 }\,.
 \label{10}
\end{multline}
In deriving expression~\eqref{10} \textit{Ansatz}~\eqref{6} has been used,
enabling us to set $\Omega_0 = \Omega(0,l) = \triangle^2(l)$. Whence it follows
that $\text{Re\,}\gamma(k) > 0$ for $k>0$, so the homogeneous state of the
chain is stable with respect to infinitely small perturbations of the particle
arrangement.

\section{Nonlinear dynamics\label{sec:2}}

\begin{figure*}
 \begin{center}
   \includegraphics[width = \textwidth]{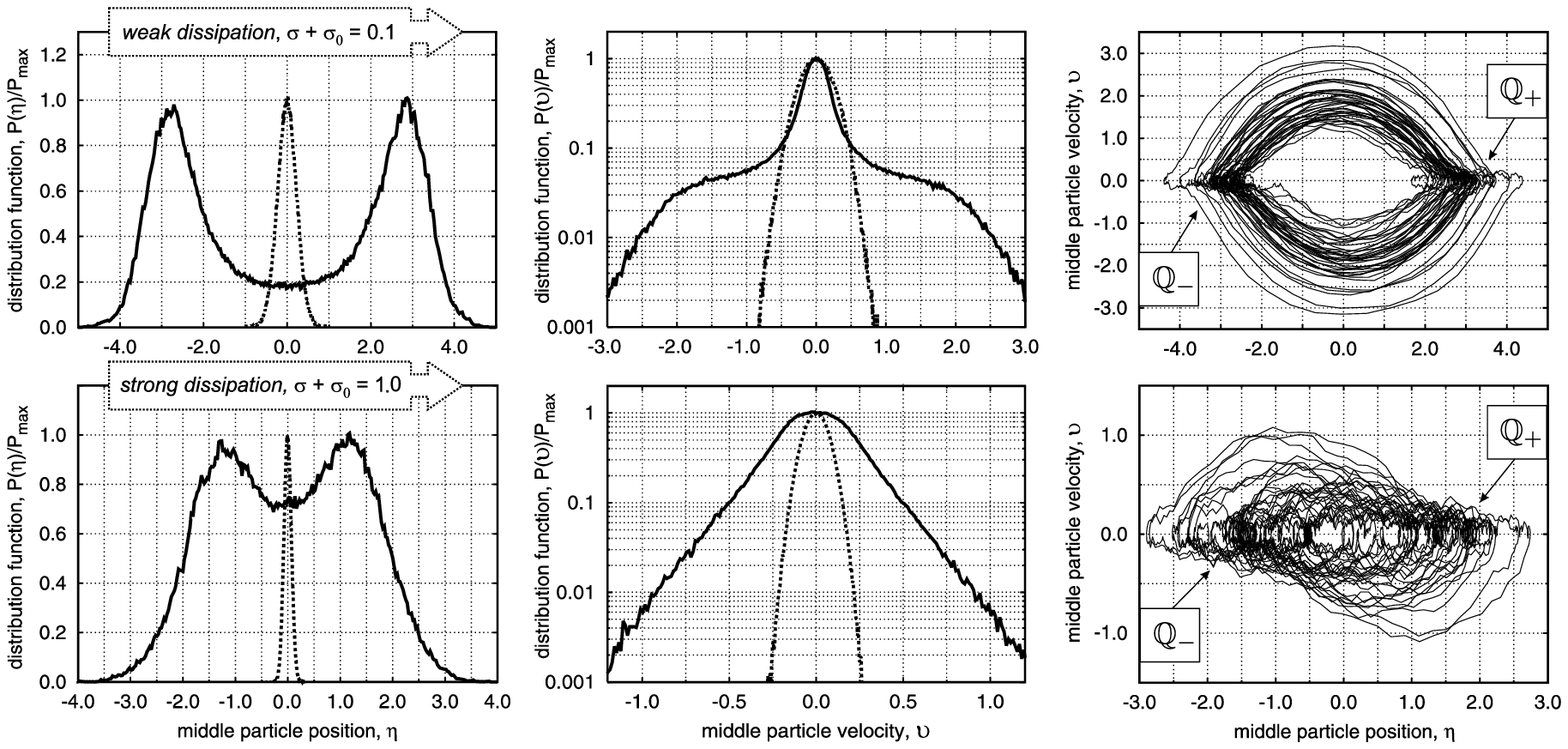}
\end{center}
\caption{
 The distribution functions of the coordinate $\eta$ and the velocity
 $\vartheta$ of the movable particle in the 3-particle ensemble. These
 distributions were obtained by averaging the simulation results over time
 interval of 100000. Solid lines correspond to the case of strong trap effect,
 $\triangle = 0.1$, dotted line match the absence of dynamical traps,
 $\triangle = 1.0$. The right widows depict some path fragment formed by the
 movable particle during time interval of 1000 time units. Other used
 parameters are $\epsilon = 0.1$ and two values of the dissipation rate $\sigma
 + \sigma_0 = 0.1$ (upper row) and $\sigma + \sigma_0 = 1.0$ (lower row).
 \label{F1p1}}
\end{figure*}

The nonlinear dynamics of the given system has been analyzed numerically. The
integration of the stochastic differential equations~\eqref{1}, \eqref{2} was
performed using the E2 high order stochastic Runge-Kutta method~\cite{RKM1}
(see also work~\cite{RKM2}). Particle collisions were implemented analyzing a
linear approximation of the system dynamics  within \textit{one} elementary
step of the numerical procedure and finding the time at which a collision has
happened. Then this step treated as a complex one was repeated. The integration
time step of 0.02 was used, the obtained results had been checked to be
actually stable with respect to decreasing the integration time step. The
ensemble of 1000 particles was studied in order to make the statistics
sufficient and to avoid a strong effect of the boundary conditions. The
integration time $T$ was chosen from 5000 to 8000 time units in order to make
calculated distributions stable. At the initial stage all the particle were
distributed uniformly in space whereas their velocities were randomly and
uniformly distributed within the unit interval.

The results of numerical simulation were used to evaluate the following partial
distributions
\begin{equation}\label{PD}
  \mathcal{P}(z) = \frac{1}{(N-2M)(T-T_0)}\sum^{N-M}_{i= M}
  \int\limits_{T_0}^{T} dt\,\delta (z - z_i(t))\,,
\end{equation}
where the time dependence $z_i(t)$ describes the dynamics of one of the
variables $\eta_i(t)$, $\vartheta_i(t)$, and $v_i(t)$ ascribed to particle $i$
and $z$ is a given point of the space $\mathbb{R}_z$ describing the symmetry
distortion $\eta$, the distortion rate $\vartheta$, and the particle velocity
$v$, respectively. The variables $\{\eta,\vartheta,v\}$ enable one to represent
the system dynamics portrait within the space
$\mathbb{R}_\eta\times\mathbb{R}_\vartheta\times\mathbb{R}_v$ or its subspace.
Besides, $N$ is the total number of particles in the ensemble and $M$ is the
number of particles located near each of its boundaries. They are excluded from
the consideration in order to weaken a possible effect of the specific boundary
conditions. The same concerns the lower boundary of time integration $T_0$, its
value is chosen to eliminate the effect of the specific initial conditions.

The numerical implementation of the integration over time in
expression~\eqref{PD} was related to the direct summation of the obtained time
series and the partition of the corresponding space $\mathbb{R}_z$ was chosen
so that the results be practically independent of the cell size. The value of
$M$ was also chosen using the result stability with respect to the double
increase in $M$. Typically the value $M\sim 50$ was chosen for $N = 1000$, for
$N = 3$, naturally, $M =1$, and $T_0\sim 500$--1000.

\subsection{Three particle ensemble}

The given oscillator chain made of three particles is actually the system
studied in part previously~\cite{we1}. In this case only the middle particle is
movable and the variables $\eta := \eta_2$ and $\vartheta := \vartheta_2$ are
its coordinate and velocity. Here we also present the results for the
3-particle ensemble in order to have a feasibility of distinguishing
characteristics of local nature from many particle effects.

Figure~\ref{F1p1} compares the distribution functions $\mathcal{P}(\eta)$ and
$\mathcal{P}(\vartheta)$ obtained in the cases where the dynamical trap effect
is absent ($\triangle = 1$) and when the dynamical traps affect the particle
motion substantially ($\triangle \ll 1$). The upper windows correspond to the
system with weak dissipation, $\sigma+\sigma_0=0.1$, whereas the lower ones are
related to the case of strong dissipation, $\sigma+\sigma_0=1.0$

\begin{figure}
\begin{center}
    \includegraphics[width=0.8\columnwidth]{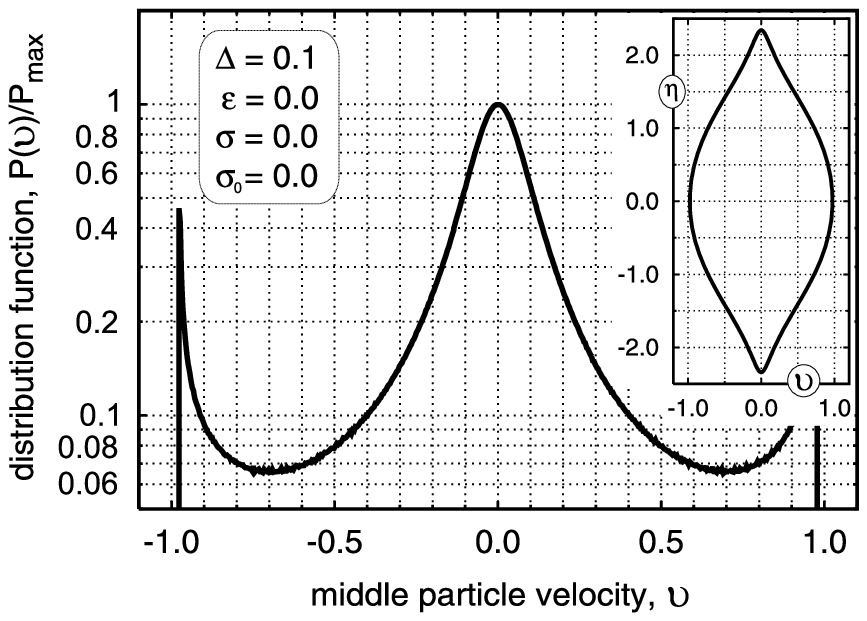}
\end{center}
\caption{
  An example of the velocity distribution $\mathcal{P}(\vartheta)$ formed by
  the movable particle of the 3-particle ensemble  without noise and
  dissipation ($\epsilon = 0$ and $\sigma = 0$). The path on the phase plane
  $\{\eta,\vartheta\}$ formed by this particle is show in the included window.
  In numerical simulation $\triangle = 0.1$ was used and a phase path of the
  velocity amplitude about unity was chosen.
  \label{F1p2}}
\end{figure}

In agreement with the previous results~\cite{we1} it is seen that the decrease
of the parameter $\triangle$, i.e. the dynamical trap intensification induces
the conversion of the function $\mathcal{P}(\eta)$ from the unimodal form to
the bimodal one, with the dissipation no more then  weakening this effect. A
new result is the essential dependence of the velocity distribution on the
dissipation rate. In the case of weak dissipation the movable particle performs
alternatively fast motions outside the trap region and slow motion inside it.
The fast motion paths connect the neighborhoods $\mathbb{Q}_{-}$,
$\mathbb{Q}_{+}$ of the $\mathcal{P}(\eta)$-function maxima, whereas the slow
motion arises when the particle wanders inside these regions. This feature is
visualized in Fig.~\ref{F1p1}, the right upper window shows a fragment of the
particle path of duration about 1000 time units. Therefore the obtained
distribution function $\mathcal{P}(\vartheta)$ as seen in Fig.~\ref{F1p1}
(middle upper window) actually is made of two monoscale components. For the
case of strong dissipation the two neighborhoods $\mathbb{Q}_{-}$ and
$\mathbb{Q}_{+}$ are not directly connected by the fast motion paths
(Fig.~\ref{F1p1}, right lower window). Now they rather uniformly spread over a
certain domain on the $\{\eta,\vartheta\}$-plane, previously, they were located
inside a sufficiently narrow layer. As a result the velocity distribution
converts into a monoscale function having a quasi-cusp form $\propto
\exp\{-|\vartheta|\}$. We relate the cusp formation to the properties of the
system dynamics near the trap region. It is justified in Fig.~\ref{F1p2}
showing the resulting velocity distribution of the movable particle when noise
and dissipation are absent. In this case the 3-particle system admits
conservation of a certain ``energy'' and the phase paths form a collection of
closed curves on the $\{\eta,\vartheta\}$-plane~\cite{we1}.

If the dynamical trap effect is absent, $\triangle = 1$, all these
distributions, as it must, are of the  Gaussian form shown in Fig.~\ref{F1p1}
with dotted lines.

\subsection{Multi-particle ensemble}

To analyze cooperative phenomena arising in such systems the dynamics of
1000-particle ensembles was implemented. Let us, first, consider local
properties exhibited by these ensembles. The term ``local'' means that the
corresponding state variable can take practically independent values when the
particle index $i$ changes by one or two. The variable $\eta_i$
(expression~\eqref{3}) may be regarded in such a manner. It describes the
symmetry of particle arrangement in space, when $\eta_i = 0$ particle $i$ takes
the middle position between the nearest neighbors, particles $i-1$ and $i+1$. A
nonzero value of $\eta_i$ denotes its deviation from this position, in other
words, a local distortion of the ensemble symmetry. The latter was the reason
for the used name of the variables $\eta_i$ as well as the variables
$\vartheta_i = d\eta_i/dt$.

\begin{figure*}
\begin{center}
 \includegraphics[width =\textwidth]{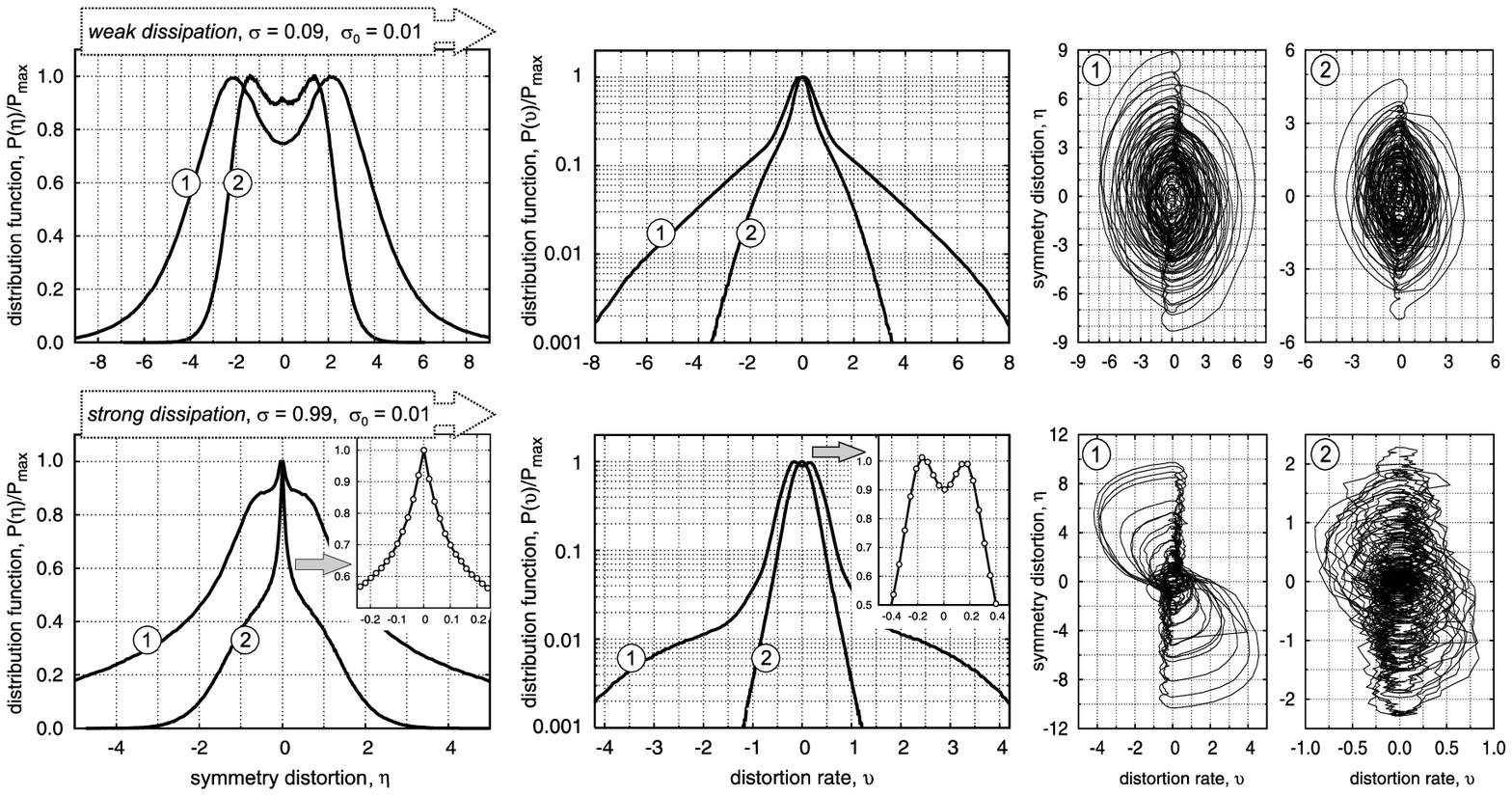}
\end{center}
\caption{
  The distribution functions of the symmetry distortion $\eta$ and the
  distortion rate $\vartheta$ for the 1000 particle ensemble with low ($l=50$,
  label 1) and high ($l=5$, label 2) density and weak ($\sigma\approx 0.1$) and
  strong ($\sigma\approx 1.0$) dissipation. The right four windows depict
  characteristic path fragments of duration of 1000 time units formed by a
  single particle with index $i=500$ on the phase plane $\{\eta,\vartheta\}$
  which was chosen due to its middle in the given ensemble. The other used
  parameters are the noise amplitude $\epsilon = 0.1$, the trap effect measure
  $\triangle = 0.1$, the small regularization friction coefficient $\sigma_0 =
  0.01$ and the regularization spatial scale $h_0 = 0.25$. The time interval
  within which the data were averaged changed from 2000 up to 5000 in order to
  make the obtained distributions stable.
  \label{Fmp1}}
\end{figure*}

Figure~\ref{Fmp1} exhibits the found distributions of the variables $\eta$ and
$\vartheta$ depending on the dissipation rate $\sigma$ and the initial distance
$l$ between the particles, i.e. their mean density. Comparison of
Fig.~\ref{F1p1} and Fig.~\ref{Fmp1} shows us that in this case of weak
dissipation the distribution functions of the symmetry distortion
$\mathcal{P}(\eta)$ and the distortion rate $\mathcal{P}(\vartheta)$ are
qualitatively similar to those of the corresponding 3-particle ensemble. Only a
few new features appear. First, for the system with high particle density
$(l=5)$ a small spike is visible at the center, $\eta = 0$, of the distribution
function $\mathcal{P}(\eta)$, which is pronounced in the case of strong
dissipation. It corresponds to the symmetrical state of the particle ensemble
being stable without dynamical traps and destroyed for the 3-particle ensemble.
In the given case ``many-particle'' effects seem to reconstruct it in part. So
in the given case the particle arrangement is characterized by three states,
two of them match the extrema of the distribution function $\mathcal{P}(\eta)$
and the symmetrical state singled out to a some degree.

As for the 3-particle ensemble the distortion rate distribution is again
composed of two monoscale components, narrow and wide ones. Previously we have
related them to the fast and slow motions. Figure~\ref{Fmp1} (upper second
window) also justifies this. The narrow component is due to the particle motion
inside the trap region and should be practically independent of the mean
distance between particles. By contrast, the wide one is to depend remarkably
on the particle density because it matches the fast motion of particles outside
the trap region and, thus, has to be affected by their relative dynamics.
Exactly this effect is demonstrated in Fig.~\ref{Fmp1} visualizing also the
corresponding properties of the particle paths.

For the 1000-particle ensemble with strong dissipation, $\sigma \approx 1.0$,
the situation changes dramatically, although the characteristic scales of the
corresponding distributions turn out to be of the same order in magnitude. In
the given case the distribution function $\mathcal{P}(\eta)$ of the symmetry
distortion has only one maximum at $\eta = 0$, however, its form is to be
characterized by two scales. In other words, it looks like a sum of two
monoscale components. One of them is sufficiently wide, its thickness is about
the same value that is obtained for the corresponding particle ensemble with
weak dissipation. Exactly this component exhibits a remarkable dependence of
the particle density, enabling us to relate it to the particle motion outside the
trap region. The other is characterized by an extremely narrow and sharp form
shown in detail in the inner window in Fig.~\ref{Fmp1} for the dense particle
ensemble. Its sharpness leads us to the assumption that ``many-particle''
effects in such systems with dynamical traps cause the symmetrical state to
be singled out from the other possible states in properties.

By contrast, the distortion rate behaves rather similar to the previous case
except for some details. When the mean particle density is high $(l=5)$ the
wide component of the distortion rate distribution disappears and only the
narrow one remains, with the latter having a quasi-cusp form
$\propto\exp\{-|\vartheta|\}$. For the system with low density the peak of the
distortion rate distribution splits into two small spikes.

These features can be explained by applying to the low row right windows in
Fig.~\ref{Fmp1}, which exhibit typical path fragments formed by motion of a
single particle on the $\{\eta\vartheta\}$-plane. Roughly speaking, now three
motion types can be singled out: some stagnation inside a narrow neighborhood
of the origin $\{\eta=0\,,\vartheta=0\}$ (visible well in the right window),
slow wandering inside the trap region that, on the average, follows a line with
a finite positive slope (clearly visible in the left window), and the fast
motion outside the trap region (visible again in the left window). The fast
motion fragments typically stem from an arbitrary point of the low motion
region and lead to a certain neighborhood of the origin. It seems that for the
system with low density particles have possibility to go sufficiently far from
the origin and during the fast motion come into the stagnation region rarely.
As a results, first, the distortion rate distribution function is of a two
scale form and contains two spikes on the peak. In the case of high density the
fast motion is depressed substantially and the system migrates mainly in the
slow motion region entering the stagnation region for many times. So the
distortion rate distribution  converts into a single-scale function and the
appearance of the symmetric state becomes often, giving rise to a significant
sharp component of the distortion distribution located near the point $\eta =
0$.

\begin{figure*}
\begin{center}
 \includegraphics[width = \textwidth]{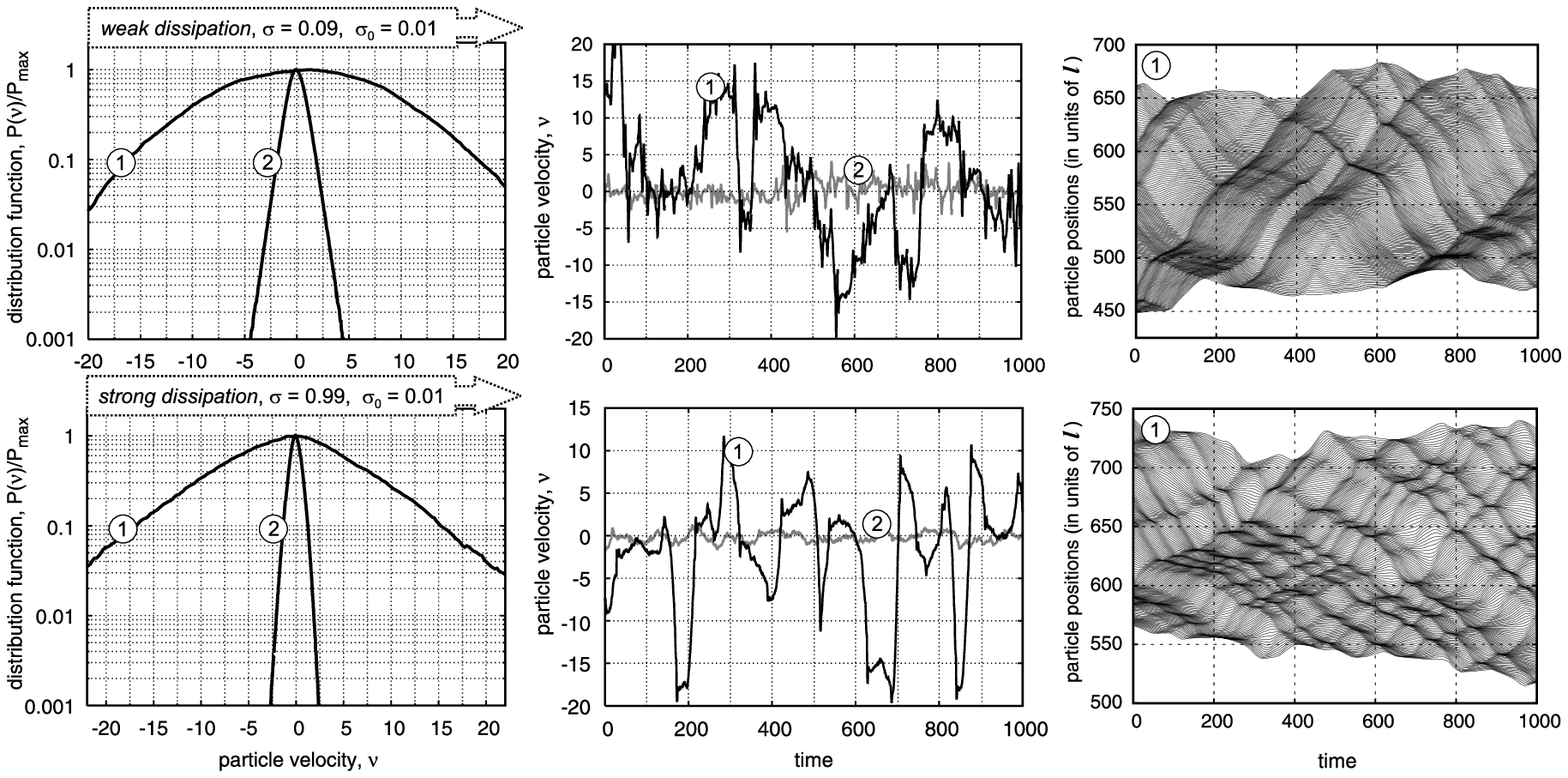}
\end{center}
\caption{
  The distribution functions of the particle velocities and the characteristic
  time patterns formed by the velocity variations of 500-th particle. Dynamics
  of the 1000-particle ensemble with low ($l=50$, label 1) and high ($l=5$,
  label 2) mean density and weak ($\sigma\approx 0.1$) and strong
  ($\sigma\approx 1.0$) dissipation was implemented for the calculation time up
  to 8000 time units to make the obtained distributions stable with respect to
  time increase. The right windows visualize the time patterns formed by 200
  paths of particle motion during 1000 time units and chosen in the middle of
  the given ensemble. The other used parameters are the noise amplitude $\epsilon =
  0.1$, the trap effect measure $\triangle = 0.1$, the small regularization
  friction coefficient $\sigma_0 = 0.01$ and the regularization spatial scale
  $h_0 = 0.25$. \label{Fmp2}}
\end{figure*}

Now we discuss the nonlocal characteristics of the 100-particle ensembles.
Figure~\ref{Fmp2} depicts the velocity distributions. As is seen it depends
essentially on both the parameters, the mean particle density and the
dissipation rate. When the mean particle density is low and the dissipation is
weak ($l = 50$ and $\sigma\approx 0.1$) the velocity distribution is
practically of the Gaussian form, however, its width gets extremely large
values about 10. We recall that without dynamical traps the width of the
corresponding distribution does not exceed 0.5 (Fig.~\ref{F1p1}). The tenfold
increase of the particle density, $l: 50\mapsto 5$, shrinks the velocity
distribution to the same order and its scale get values similar to that of the
distortion rate distribution in magnitude. However in this case the form of the
velocity distribution is a monoscale function of the well pronounced cusp form
$\propto \exp\{-|v|\}$. In the case of strong dissipation ($\sigma\approx 1.0$)
the situation is opposite. The system with low density ($l=50$), as previously,
is characterized by an extremely wide velocity distribution, its width is about
10. However, now its form deviates substantially from the Gaussian one. For the
corresponding ensemble with high density ($l=5$) the velocity distribution is
Gaussian with width about 1. The latter, nevertheless, is much larger then the
same width in the absence of dynamical traps.

These features of the velocity distribution characterizes the cooperative
behavior of particles rather then their individual dynamics. In other words,
there should be strong correlations in the motion of not only neighboring
particles but also distant ones. Therefore the velocity variations responsible
for the formation of such distributions describe in  fact the motion of
mutliparticle clusters. To justify this we apply to the middle column windows
in Fig.~\ref{Fmp2}. They visualize some typical fragments of the time patterns
formed by the velocities of individual particles. When the mean particle
density is low ($l = 50$), these patterns look like a sequence of fragments
$\{v_\alpha\}$ inside which the particle velocity varies in the vicinity of
some level $v_\alpha$. The values $\{v_\alpha\}$ are rather randomly
distributed inside a certain region of thickness $V\sim 10$ in the vicinity of
$v= 0$. The continuous transitions between these fragments occur via sharp
jumps. The typical duration of these fragments is about $T\sim 100$, which
enables us to regard them as long-lived states because the temporal scales of
individual particle dynamics are about several units. Moreover, these
long-lived states can persist if only a group of many particles moves as a
whole because the characteristic distance $L$ individually traveled by a
particle involved into such state is about $L\sim VT\sim 1000\gg l$.

The spatial structure of these cooperative states is visualized in
Fig.~\ref{Fmp2}, right column windows. They depict time patterns formed by
paths $\{x_i(t)\}$ of 200 particles of duration about 1000 time units. These
particles were chosen in the middle part of the 1000-particle ensembles with
low density. For high density ensembles such patterns also develop but are not
so pronounced. As is seen a large number of different mesoscopic states formed
in these systems. They differ from one another in size, the direction of
motion, the speed, the life time, \textit{ets}. Moreover, the life time of such
a state can be much longer then the characteristic time interval during which
particles forming it currently will belong to this state individually. Besides,
the found patterns could be classified as hierarchical structures. Some
relatively small domains formed by cooperative motion of individual particles
in their turn make up together larger superstructures. In other words, the
observed long-lived cooperative states have their ``own'' life independent, in
some sense, of the individual particle dynamics. The latter properties are the
reason for regarding them as certain dynamical phases arising in the systems
under consideration  due to the dynamical traps affecting the individual
particle motion. The term ``dynamical'' has been used to underline that the
complex cooperative motion of particles is responsible for these long-lived
states, without the continuous particle motion such states cannot exist.

The obtained results are summarized in the following section.

\section{Conclusion}

A rather simple model of an oscillator chain, a one-dimen\-sional particle
ensemble, with dynamical traps and additive white noise has been considered. It
should be noted that the regular ``force'' governing the individual dynamics of
particles has no stationary points except for one matching the system
homogeneous state. The latter is locally stable for all the possible values of
the system parameters. Nevertheless, as has been demonstrated numerically, the
sufficiently strong dynamical trap effect accompanied with white noise gives
rise to a wide variety of anomalous and cooperative phenomena.

In particular, the local symmetry of particle arrangement (described by
variables~\eqref{3} called the symmetry distortion) can exhibit kinetic phase
transitions. Depending on the mean particle density and the dissipation rate
the distribution function of the symmetry distortion takes either a bimodal
form or a twoscale unimodal form, with the latter possessing extremely sharp
spike (Fig.~\ref{Fmp1}). The distortion rate distribution also either is
characterized by two scales or is of the cusp form
$\propto\exp\{-|\vartheta|\}$ smoothed, naturally, inside a narrow transition
region.

The cooperative phenomena arising in these system have been studied analyzing
the velocity distributions and visualizing some time pattern formed by the
particle dynamics. When the mean particle density is sufficiently low the
velocity distributions are characterized by the extremely large widths. For the
system with high density and low dissipation the velocity distribution function
takes also a quasi-cusp form. The visualized time pattern of the velocity
dynamics of a single particle has demonstrated the presence of the long-lived
states. They look like a sequence of fragments $\{v_\alpha\}$ within which the
particle velocity varies in the vicinity of some level $v_\alpha$ continuously
joined by sharp jumps in the particle velocity. The velocity levels are rather
uniformly distributed inside a wide interval and the life time of these
fragments exceeds essentially the time scales of the individual dynamics of
particles. It has been shown that such long-lived states can persist if only
multiparticle clusters moving as a whole are formed.

The visualized patterns made up of paths of 200 particles have shown also the
presence of such cooperative structures. Moreover it has become clear that
these long-lived states persist independently in some sense of the individual
dynamics of particles forming them currently. In other words, the life time of
such a state can exceed substantially the time interval during which the
particles forming it at a current time belong to it.  Keeping the latter in
mind we refer to them as to dynamical states. These states in turn can form
superstructures, so the observed patterns are classified as hierarchical
structures.

\begin{acknowledgement}

This work was supported in part by DFG Project MA 1508/6, RFBR Grant
02-02-16537, Grant B0056 of Russian Program ``Integration'', and Moscow
Government Grant 1.1.133.

\end{acknowledgement}

\end{document}